\documentclass[12pt,preprint]{aastex}

\slugcomment{} \shorttitle{Heating Processes Diagnostics of Solar
Flares} \shortauthors{Cheng et al.}
\begin{document}
\title{Diagnostics of the Heating Processes in Solar Flares Using Chromospheric Spectral Lines}
\author{J. X. Cheng, M. D. Ding and J. P. Li}
\affil{Department of Astronomy, Nanjing University, Nanjing
210093, China}
\begin{abstract}
We have calculated the H$\alpha$ and Ca {\sc ii} 8542 {\AA} line
profiles based on four different atmospheric models, including the
effects of nonthermal electron beams with various energy fluxes.
These two lines have different responses to thermal and nonthermal
effects, and can be used to diagnose the thermal and nonthermal
heating processes. We apply our method to an X-class flare that
occurred on 2001 October 19.  We are able to identify
quantitatively the heating effects during the flare eruption. We
find that the
 nonthermal effects at the outer edge of the flare ribbon are more
notable than that at the inner edge, while the temperature at the
inner edge seems higher. On the other hand, the results show
 that nonthermal effects increase
rapidly in the rise phase and decrease quickly in the decay phase,
but the atmospheric temperature can still keep relatively high for
some time after getting to its maximum. For the two kernels that
we analyze, the maximum energy fluxes of the electron beams are
$\sim$ 10$^{10}$ and 10$^{11}$ ergs cm$^{-2}$ s$^{-1}$,
respectively. However, the atmospheric temperatures are not so
high, i.e., lower than or slightly higher than that of the weak
flare model F1 at the two kernels. We discuss the implications of
the results for two-ribbon flare models.
\end{abstract}
\keywords{Sun: flares --- Sun: chromosphere}
\section{Introduction}
Solar flares are one of the most significant active phenomena in
the solar atmosphere. At present the widely accepted flare model
based on magnetic reconnection is the CSHKP model, which was
developed by \citet{car64}, \citet{Stu66}, \citet{hira74} and
\citet{kopp76}. The model assumes that pre-existing closed
magnetic loops in the corona are torn open by the force of the
filament eruption as a result of magnetic instability.
Subsequently, a current sheet is stretched out and energy stored
in the magnetic field is released at the reconnection point. As a
consequence, a cusp-shaped loop structure appears which was
observed by {\it Yohkoh} in soft X-rays  \citep{tsuneta92} and by
$Reuven$ $Ramaty$ $High$ $Energy$ $Solar$ $Spectroscopic$ $Imager$
$(RHESSI)$ in hard X-rays (HXRs) \citep{sui03}. As successive
magnetic reconnection goes on, the loop top HXR source
\citep{masuda94} rises \citep{sui04,liu04} and the footpoint
sources  separate from each other
 \citep{fletcher01,liu04}. Recent observations reveal
 a more HXR source above the flare loop, called the coronal source \citep{sui04}.
 The temperature in between the loop top and the coronal source is found to be
 higher than that in  lower and higher altitudes; therefore, this place is
 regarded as associated with the formation and development of a current sheet.

It is known that the primary energy release during a solar flare
 (the current sheet) is in the corona. The released energy results in a bulk heating
  of the plasma and acceleration of charged particles. Therefore, the chromosphere
  can be heated by either a conduction front or electron beam bombardment \citep{brown73,can74,brown78,ems78}. Other energy transport processes, such as heating
by an energetic proton beam \citep{lin76,ems83,hen93} and soft X-ray irradiation \citep{hen77,mac78,gan90,ber04} have also been
invoked. Many studies have been devoted to the spatial
distribution and temporal evolution of the thermal/nonthermal
 heating signatures. \citet{can93} argued that energetic electrons favor to
occur at the edge of high vertical currents. This is confirmed by \citet{masuda01}
 that the spectrum tends to be
harder at the outer edge of a ribbon than in the inner edge.
\citet{czay99} found that strong upflows, revealed by the blue
shifts in EUV lines, appear at the outer edges of flare ribbons.
\citet{li04} showed that chromospheric downflows, revealed by the
red asymmetries in  the H$\alpha$ lines, are also the most obvious
at the outer edges. These findings are basically consistent with
the two-ribbon flare model, in which the outer edges map the
footpoints of newly reconnected flare loops. The general scenario
of flare evolution implies that the thermal/nonthermal heating
processes can vary both spatially and temporally. Therefore, it is
interesting to devise a method to diagnose the relative
importances of thermal and nonthermal heating processes in a
flare.

The purpose of this paper is to use two different chromospheric
lines, H$\alpha$ and Ca {\sc ii} 8542 {\AA} lines, to diagnose the
heating processes in flares. The H$\alpha$ line is the most
observed and studied chromospheric line in solar flare
spectroscopy. Theoretical calculations have shown how the
H$\alpha$ line varies with different flare parameters like the
nonthermal electron flux, conduction flux, and coronal pressure
\citep{ric83,can84}. In particular, when considering the
nonthermal excitation and ionization effects by electron beam, the
H$\alpha$ line can be enhanced significantly \citep{fang93,kav02}
. On the other hand, the Ca {\sc ii} 8542 {\AA} line is less
sensitive to the nonthermal effects, though also enhanced to some
extent. This line is more influenced by the coronal pressure and
chromospheric temperature. The different responses of the
H$\alpha$ and Ca {\sc ii} 8542 {\AA} lines to thermal/nonthermal
effects make it possible to diagnose the heating processes using
these two lines together.

We make non-LTE calculations of the H$\alpha$ and Ca {\sc ii} 8542
{\AA} line intensities for different atmospheric models and
nonthermal electron beams. The results are used to diagnose the
processes of a flare on 2001 October 19 in detail. The paper is
organized as follows.
 The method for model
calculations is given in \S2. Theoretical results are presented in
\S3. \S4 shows the diagnostics of the flare, followed by
discussions and conclusions in \S5.

\section{Method of model calculations}
The flare eruption is a very complicated process in the solar
atmosphere. It involves a drastic change of the atmospheric
conditions subject to a time-varying energy input. Semi-empirical
models have been widely used to reproduce the observed flare
spectra (e.g. \citet{machado80}; \citet{gan87}). For simplicity,
we adopt four atmospheric models to represent the atmospheric
status at different phases of a flare. The temperature
distributions versus column mass density of the four models are
showed in Figure 1. FQ is the model for the quiet-Sun
\citep{vernazza81} which can be regarded as the preflare status.
F1 and F2 are weak and strong flare models, respectively
\citep{machado80}. FA is an interpolation between F1 and F2.
Therefore,  the sequence FQ-F1-FA-F2 represents roughly the
variation behavior of a flare atmosphere. In general, from FQ to
F2, the chromospheric temperature becomes higher, the transition
region tends to be lower, and the coronal pressure
 increases. The temperature in the photosphere has almost no change in these models.

On the other hand, the role of the electron beam should be taken
into account in the model calculations. Considering the variation
of the electron beam as revealed by the HXR observations, we adopt
five different energy fluxes 0, 10$^{9}$, 10$^{10}$, 10$^{11}$,
and 10$^{12}$ ergs cm$^{-2}$ s$^{-1}$ for each model. We assume a
power law distribution for the electron beam with a spectral index
$\delta=4$ and low-energy cutoff $E_c$=20 keV. In fact, the
calculated results are not affected much by the latter two
parameters.

Therefore, we make  non-LTE calculations based on the atmospheric
models with different electron beams to get the line profiles of
H$\alpha$ and Ca {\sc ii} 8542 {\AA}. We include the nonthermal
excitation and ionization effects by the electron beam in the
calculations. Each model and each electron beam yield a specific
set of the H$\alpha$ and Ca {\sc ii} 8542  {\AA} profiles.
\section{Theoretical line profiles}

Figures 2 and 3 show the theoretical profiles of H$\alpha$ and Ca
{\sc ii} 8542 \AA, respectively. In each panel, the profiles are
for the same atmospheric model but different electron beam fluxes.
It can be seen that the H$\alpha$ and Ca {\sc ii} 8542 {\AA} lines
have quite different responses to the model parameters, in
particular, the electron beam flux. As the  flux increases, the
H$\alpha$ line intensity is enhanced greatly and the profile
becomes broadened; at the same time, the central reversal of the
H$\alpha$ profile becomes more
 obvious. These features are typical characteristics of hydrogen
Balmer lines  under the circumstance of nonthermal heating.
However, the Ca {\sc ii} 8542 {\AA} line is less sensitive to the
nonthermal electron beam especially for fluxes higher than
10$^{10}$ ergs cm$^{-2}$ s$^{-1}$. Considering these facts, we
find a convenient parameter that can be used to reflect the
different responses of the two lines to the nonthermal effects,
that is, the wavelength-integrated intensity as described by
\begin{equation}
EW =
\int^{\lambda_{c}}_{-\lambda_{c}}\frac{I_{\lambda}-I_{\lambda_0}}{I_{c}}d\lambda
\end{equation}
In equation (1), $I_{c}$ is the continuum intensity, and
$I_{\lambda}$ and $I_{\lambda_0}$ are the line intensities for the
flare and the quiet-Sun, respectively. As for the integration
range, we adopt $\lambda_{c}=6$ {\AA} for H$\alpha$ and 1 {\AA}
for Ca {\sc ii} 8542 {\AA}, respectively. The merit of using  such
a wavelength-integrated quantity is that this value is independent
of the macro-turbulent velocity, which can affect the profiles
greatly. In fact, the parameter defined by equation (1) is similar
to the equivalent width of lines. Therefore, we will use the term
``equivalent width (EW)" instead of ``wavelength-integrated
intensity" hereafter.

We note that in the cases of non-quiet models, the far wings of
line profiles are slightly above zero after subtraction of the
quiet profile. This is not the extension of broad wing emission,
but just reflects a very small continuum enhancement in these
models relative to the quiet-Sun.

Figure 4 shows the equivalent width of Ca {\sc ii} 8542 {\AA}
against that of H$\alpha$. Each solid
 line refers to the same atmospheric model. The five asterisks from
 left to right denote different nonthermal electron beam fluxes of 0, 10$^{9}$, 10$^{10}$, 10$^{11}$, and
10$^{12}$ ergs cm$^{-2}$ s$^{-1}$, respectively. The curve shows
that the equivalent width of H$\alpha$ increases very quickly with
increasing nonthermal effects. However, the Ca {\sc ii} 8542 {\AA}
equivalent width is more sensitive to the thermal models when the
flux is higher than 10$^{10}$ ergs cm$^{-2}$ s$^{-1}$. We can
superimpose the observational points (the observed set of Ca {\sc
ii} 8542 {\AA} versus H$\alpha$ equivalent widths) on this EW-EW
plot. The trajectory of the observational points reflects the time
evolution of the model atmosphere and the electron beam. From
this, we can also judge the relative importances of the thermal
and nonthermal heating processes. For example, if the trajectory
is mostly along the solid lines, then the nonthermal effects
dominate; otherwise, if the trajectory is across the solid lines,
then the thermal heating is more important. For most cases, both
of these two effects work together
 during the flare evolution.

There are some limitations of the above method to diagnose the
flare processes. First, the heating mechanisms during solar flares
are so complicated; there are some other factors than the effects
of temperature rise and nonthermal electron beam, such as the
return current (see \S5),
 that can
influence the line intensities. Second, we use a fixed spectral
index and low-energy cutoff that can in fact vary during the
flare. Third, using of the equivalent width sacrifices some useful
characteristics like the shift and asymmetry of the line profile.
However, owing to the lack of our knowledge of some physical
parameters, in particular the macro-turbulent velocity, such a
method is still a simple and practical method to diagnose
thermal/nonthermal processes in flares.

\section{Diagnostics to the 2001 October 19 flare}
\subsection{Observation}
A two ribbon flare occurred in NOAA Active Region 9661
(N16$^{\circ}$, W18$^{\circ}$) on 2001 October 19. According to
the $Solar$ $Geophysical$ $Data$, it is an X1.6/2B class flare
associated with a CME event. The flare lasted from 00:47 UT to
01:13 UT, reaching its maximum at 01:05 UT.  Observations of
H$\alpha$ and Ca {\sc ii} 8542 {\AA} line profiles were made by
the Solar Tower of Nanjing University \citep{huang95,ding99}. The
flare was well observed from the beginning to the end. An analysis
of the multi-wavelength data has been done by \citet{li04}. The
results can be summarized as follows. The maximum velocity seems
to be located at the outer edges of the flare ribbons. The flare
ribbons contain four H$\alpha$ kernels denoted as K1, K2, K3, and
K4 (Fig. 5). They  are associated with two hard HXR peaks,
respectively (Fig. 6). Kernels K1 and K2 correspond to the first
peak at 00:55:23 UT while kernels K3 and K4 correspond to the
second peak at 01:00:06 UT. In this paper, we select kernels K2
and K3 for study. They belong to two different flare loops. Kernel
K2 is associated with a weak HXR source while K3 is related to a
strong HXR source. The relative importance   of thermal conduction
and nonthermal electron beam may therefore differ in the two
kernels.

\subsection{Temporal evolution}
We extract the observed line profiles at the brightest points in
kernels K2 and K3. Since the brightest point can vary in space
with time, we fix it to be the one during the flare maximum time.
A time series of line profiles are thus obtained for each kernel.
The observed equivalent width is calculated in the same way as the
theoretical one. We can then compare the observations with
theoretical calculations.

 Figure 7 shows the time evolution of kernel K2 in the EW-EW plot
  from 00:51:26 UT to 00:56:18 UT. In the rise phase of the flare,
  the temperature of the atmosphere increases until the
 maximum phase. At the same time, the nonthermal electron beam flux
 varies from 0 to greater than 10$^{10}$ ergs cm$^{-2}$ s$^{-1}$.
 During the gradual phase, the trajectory of the observational points goes back nearly along the initial
 path, only that the nonthermal electron beam is weaker and the
 temperature is a little higher than in the rise phase. In general, these
 two effects are not so strong in kernel K2. This is consistent
 with the fact that kernel K2 corresponds to a weak HXR source.

 Figure 8 shows the temporal evolution of kernel K3 from 00:53:49 UT to 01:45:56 UT. It is associated
 with a strong HXR source. During the rise phase, the nonthermal
 electron flux increases very rapidly and reaches a quite high
 value before the maximum phase. The maximum flux is even
 greater than 10$^{11}$ ergs cm$^{-2}$ s$^{-1}$. At
 the same time, the temperature is also enhanced but not so
 obvious considering the very strong electron beam. The temperature is only a little higher than that of the F1
 model but lower than that of the FA model.
  In the gradual phase, the nonthermal electron flux decreases
  quickly;
  however, the atmosphere can keep hot for some time. Compared to kernel K2 that is associated with a weak
  HXR source, the nonthermal electron flux at kernel K3 is much larger.
  This is consistent with HXR observations.
\subsection{Spatial distribution}
To check the spatial distribution of the heating signatures, we
draw a line across the flare ribbon containing kernel K3.
Twenty points are selected along the line with a uniform
space.
 The trajectory of these observational points from the inner edge to the
outer edge in the EW-EW plot is shown in Figure 9. From the inner
to outer edges, the nonthermal effects become more and more
obvious. Nonthermal electron flux varies from less than 10$^{10}$
ergs cm$^{-2}$ s$^{-1}$ to greater than 10$^{11}$ ergs cm$^{-2}$
s$^{-1}$. The position where the nonthermal effect gets to its
maximum is denoted by a solid circle superimposed on the line in
Figure 5. Obviously, it is located at the outer edge of the flare
ribbon. Generally speaking, the nonthermal effects at the outer
edge of the flare ribbon are more significant than at the inner
part. On the contrary,  the atmospheric temperature is higher at
the inner edge than at the outer edge.

 \section{Discussions and conclusions}
In general, the heating mechanisms during a solar flare are
complicated, i.e., a combination of different processes. Thermal
conduction and nonthermal  electron beam bombardment are the two
mostly discussed and important energy transport processes. Many
studies are focused on solving this controversial and undetermined
problem. \citet{Saint05} deduced the thermal and
nonthermal energies for a flare and found that they are of the
same magnitude. However, in some other flares or in a specific
phase or specific region of flares, it is possible that either
thermal heating or nonthermal heating is dominant (e.g. \citet{li05}; \citet{ji04}). If seen  in chromospheric lines, the
thermal component in the light curve is thought to be delayed by
some time compared to the nonthermal component. This time delay is
case-dependent.  Usually we cannot easily separate the thermal and
nonthermal components from the light curve only. The detailed
spectra provide a diagnostic tool to distinguish these two
components.

 Previous studies have shown that the nonthermal
excitation and ionization effects caused by an electron beam
bombardment have a great influence on chromospheric line profiles
\citep{fang93}. Since the H$\alpha$ and Ca {\sc ii} 8542 {\AA}
lines have different responses to thermal and nonthermal processes
as discussed above,  we can diagnose different heating mechanisms
using the observations of the two lines. In this work, we simply
use the equivalent widths of the lines, which refrains us from
invoking some unclear parameters like the macro-turbulent
velocity. Of course, the HXR emission is closely related to
nonthermal electrons; however, the thermal heating effects can
better be studied through checking chromospheric lines.  In our
analysis, we choose two lines, one of which is sensitive to
thermal effects while another is sensitive to nonthermal effects,
to diagnose these two heating processes. This method is tested to
be practical and useful. Recently, \citet{kar04} argued that the
return current collision and excitation could significantly
enhance the line radiation. If considering this effect, the
deduced flux of the nonthermal electron beam should be somewhat
lower. However, this does not influence our main conclusions.

The spatial distribution and temporal evolution of thermal and
nonthermal effects during a solar flare are also complex.
\citet{li04} concluded that the  chromospheric downflow velocity
tends to appear at the outer edge of  flare ribbons. This
indicates that the nonthermal effects at the outer edge of flare
ribbons are  the most significant. In the current work, by using a
novel and different method, we also find that in the 2001 October
19 flare, the nonthermal effects at the outer edge are more
distinct than that at the inner edge of flare ribbons. However,
the chromospheric temperature at the inner edge seems higher than
that at the outer part. The above results support the general
scenario of flare development: successive magnetic reconnection
occurs to form new flare loops whose footpoints are shown as the
outer edge of flare ribbons. It is conceivable that in the newly
formed loops, the nonthermal effects are more obvious than in the
old loops. At the two kernels K2 and K3 in the 2001 October 19
flare, the electron beam fluxes increase quickly in the initial
phase and decrease rapidly in the gradual phase while the thermal
effects change gradually.

\acknowledgments We thank the referee for valuable comments that
helped to improve the paper. This work was supported by the
Scientific Research Foundation of Graduate School of Nanjing
University, FANEDD under grant 200226, NSFC under grants 10403003,
10221001, and 10333040, and NKBRSF under grant 2006CB806302.

\clearpage
\begin{figure}
\epsscale{1.0} \plotone{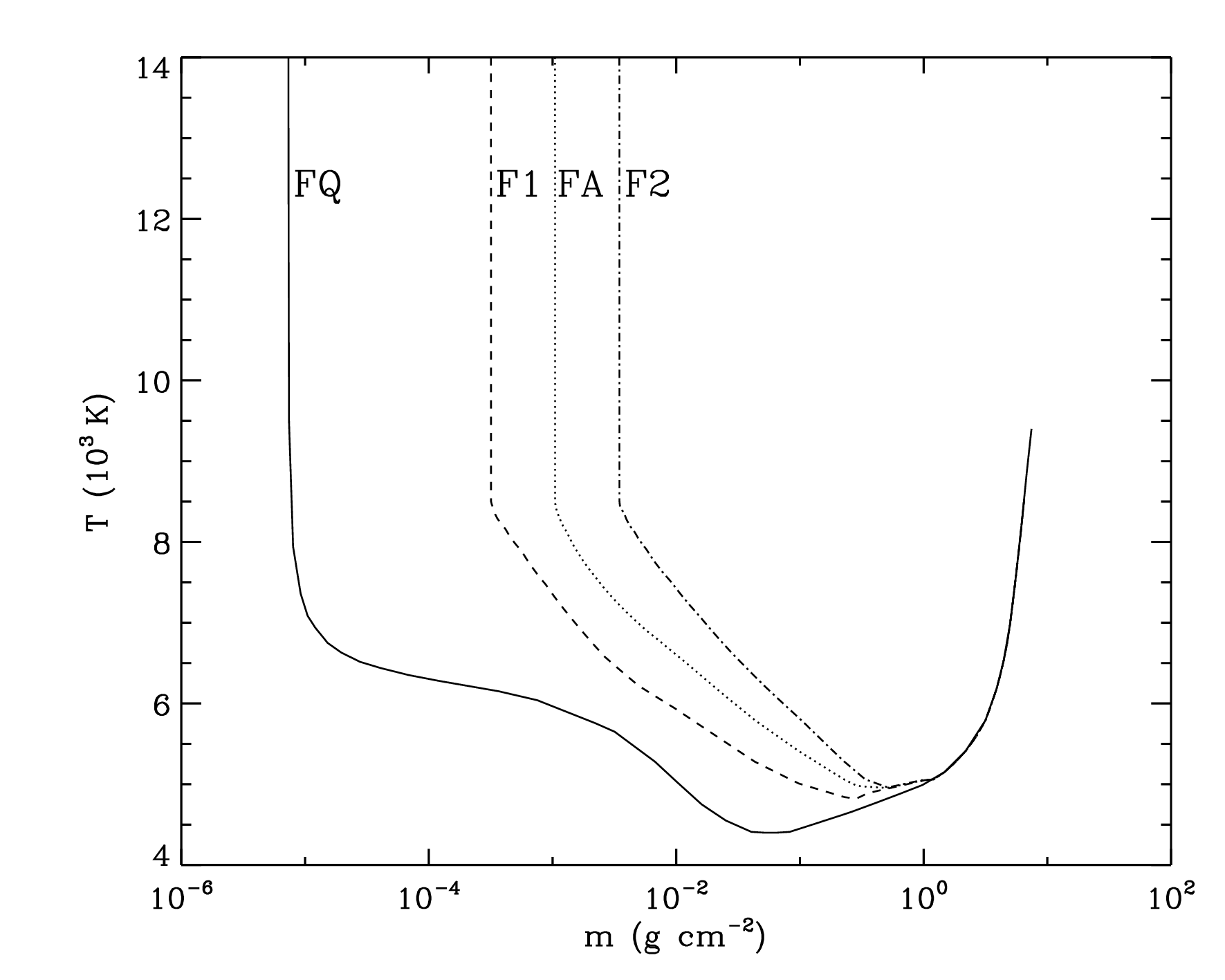} \caption{Four atmospheric models
adopted in calculations: quiet-Sun model FQ (Vernezza et al.
1981), weak flare model F1 and strong flare model F2 (Machado et
al. 1980). Model FA is an interpolation between F1 and F2.}
\end{figure}
\begin{figure}
\epsscale{1.0} \plotone{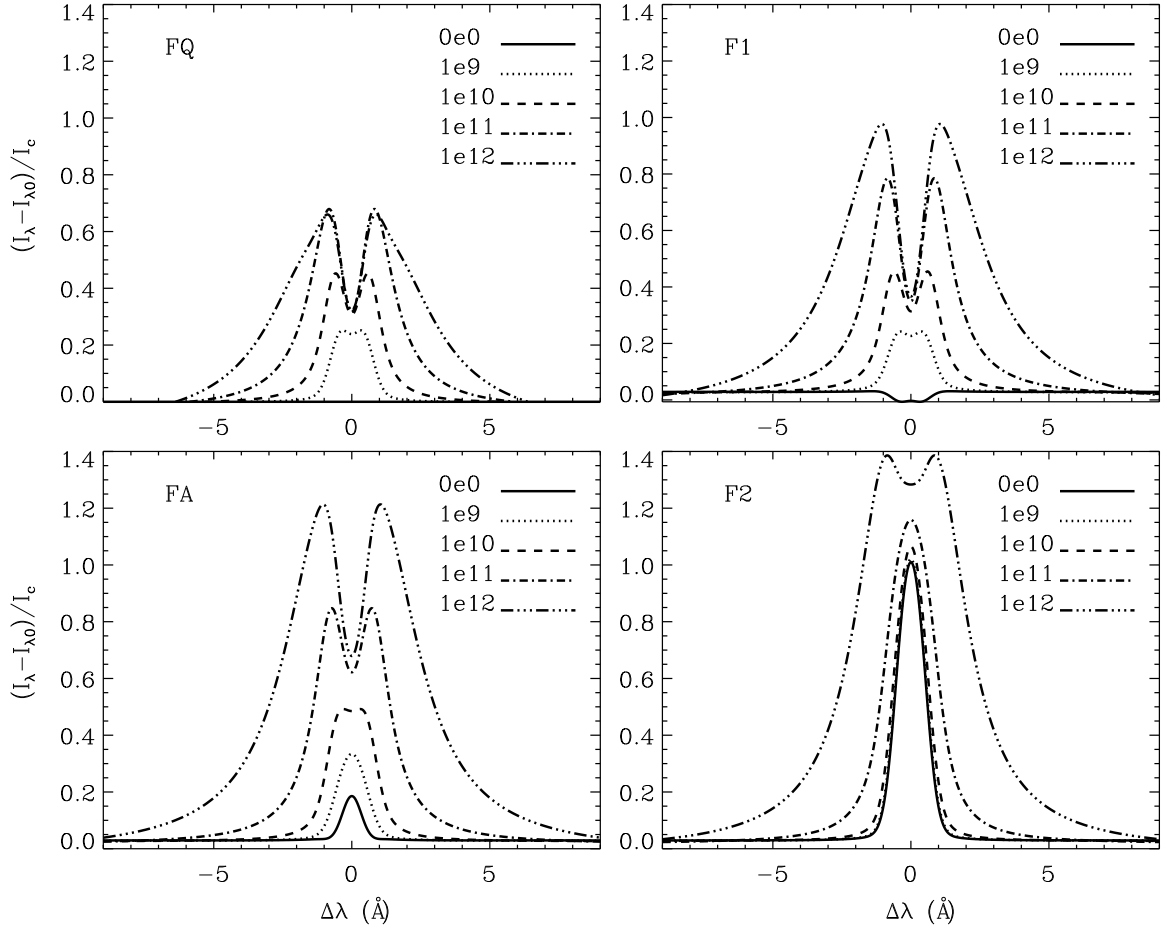} \caption{H$\alpha$ line
profiles calculated from the four atmospheric models bombarded by
electron beams with various energy fluxes. }
\end{figure}
\begin{figure}
\epsscale{1.0} \plotone{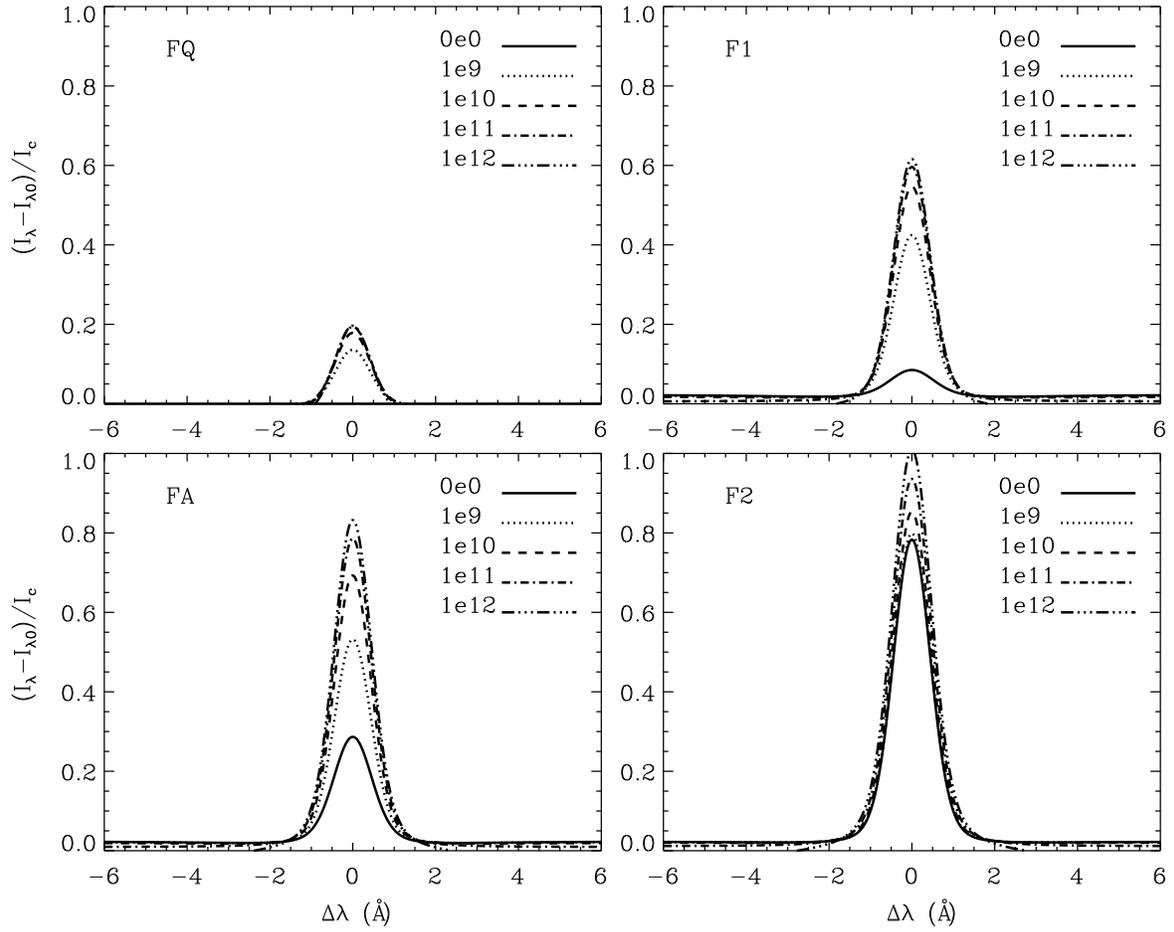} \caption{Same as Fig. 2, but for
the Ca {\sc ii} 8542 {\AA} line profiles. }
\end{figure}
\begin{figure}
\epsscale{1.0} \plotone{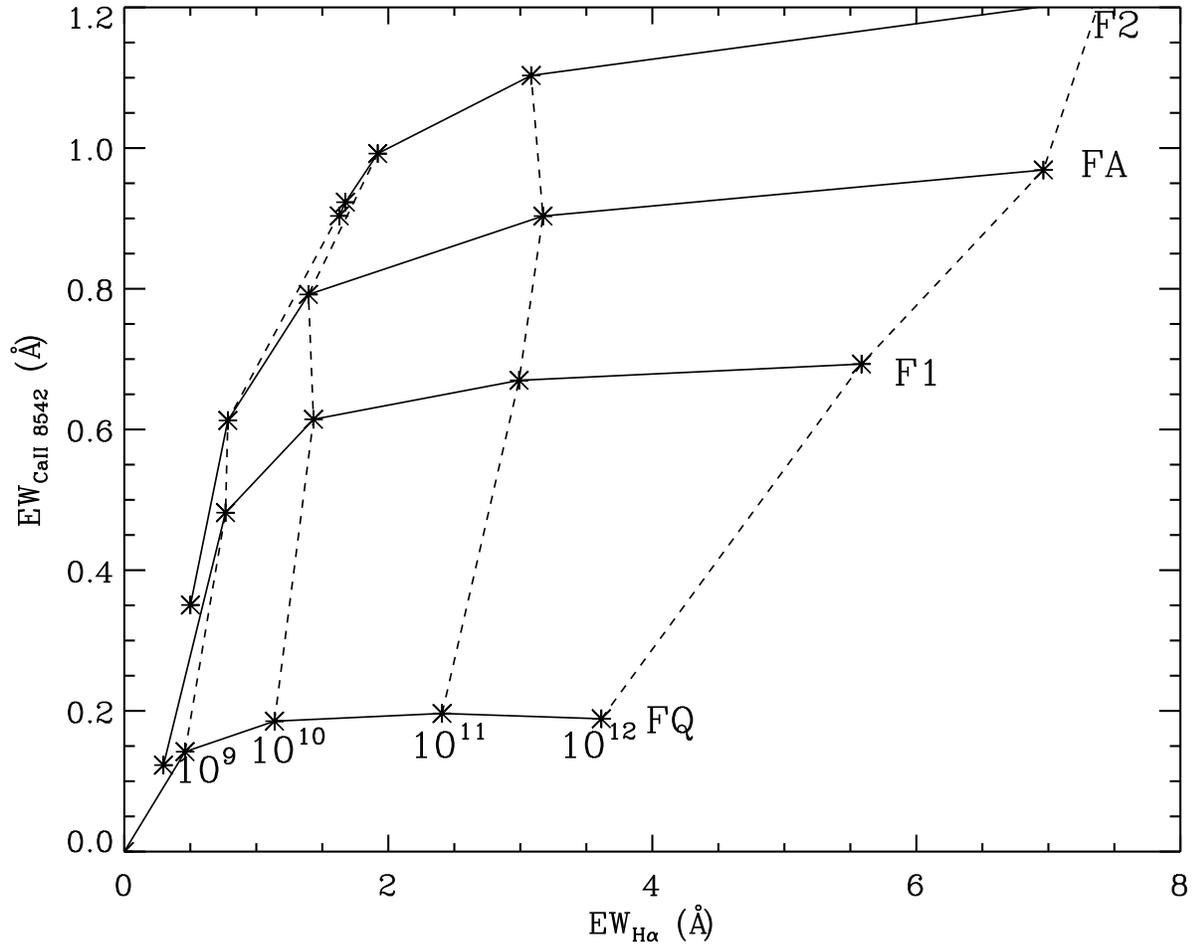} \caption{Equivalent width of Ca
{\sc ii} 8542 {\AA} versus that of H$\alpha$. Each solid curve is
for the same atmospheric model, in which the points from left to
right refer to different electron beams shown under the bottom
curve.}
\end{figure}

\begin{figure}
\epsscale{1.} \plotone{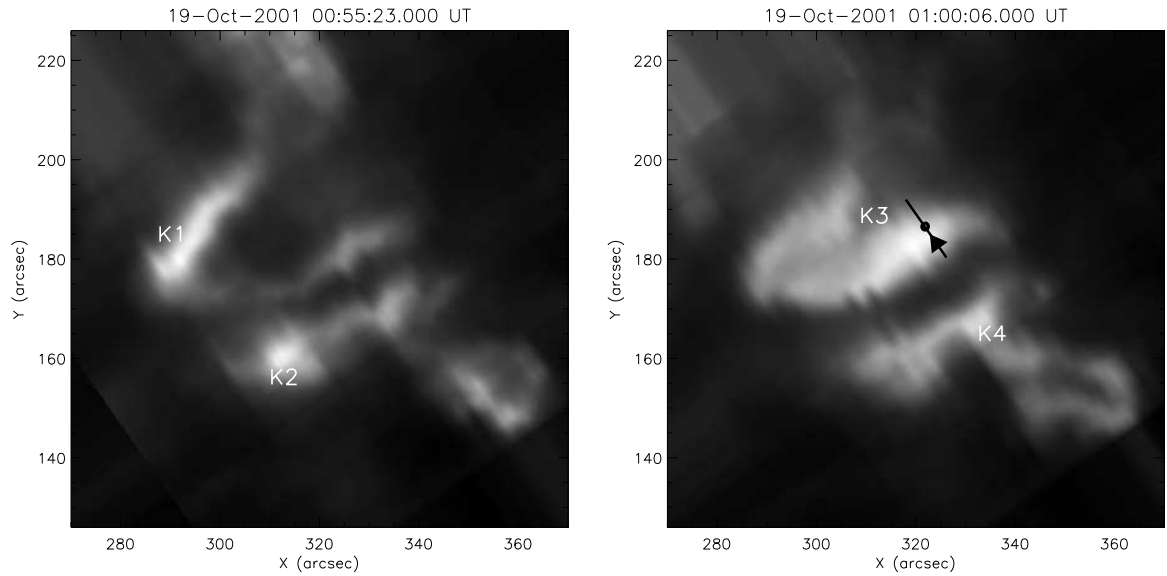} \caption{H$\alpha$ line-center
images observed by the Solar Tower of Nanjing University at
00:55:23 UT (left) and 01:00:06 UT (right) on 2001 October 19,
reconstructed from the two-dimensional H$\alpha$ spectra. The
field of view is $100''$$\times$$100''$. North is up and east is
to the left. The line across the flare ribbon is used to show the
spatial variation of the Ca {\sc ii} 8542 {\AA} versus H$\alpha$
equivalent widths in Fig. 9. The solid circle superimposed on the
line indicates the position of maximum nonthermal effect. The
arrow denotes the direction from inner to outer edges of the
ribbon.}
\end{figure}

\begin{figure}
\epsscale{1.} \plotone{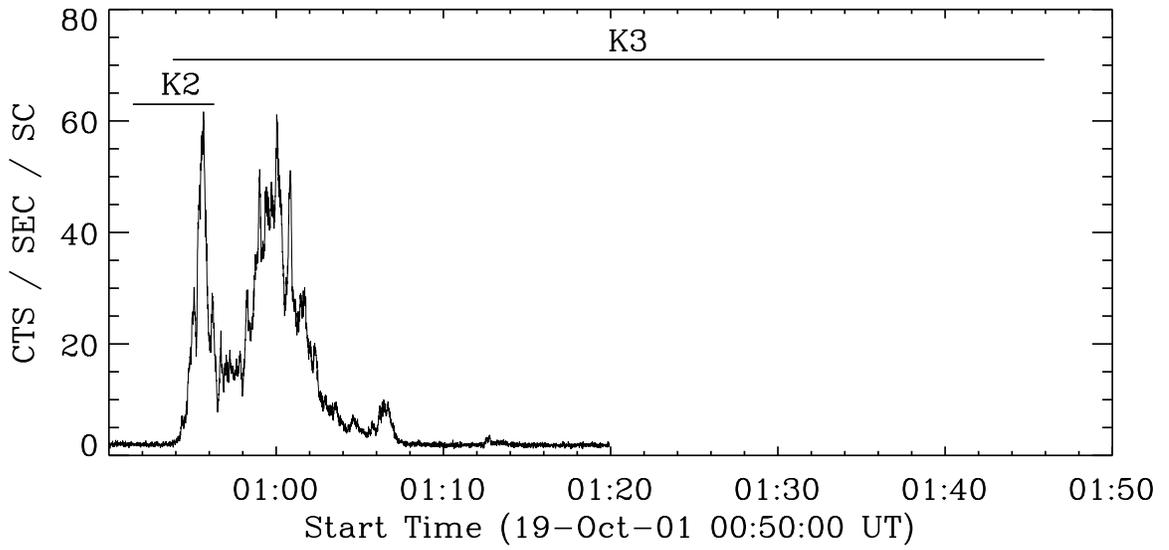} \caption{Temporal evolution of HXR
flux in the 33-53 keV band observed by {\it Yohkoh}. The
horizontal lines mark the time ranges of observational points for
K2 (Fig. 7) and for K3 (Fig. 8).}
\end{figure}

\begin{figure}
\epsscale{1.0} \plotone{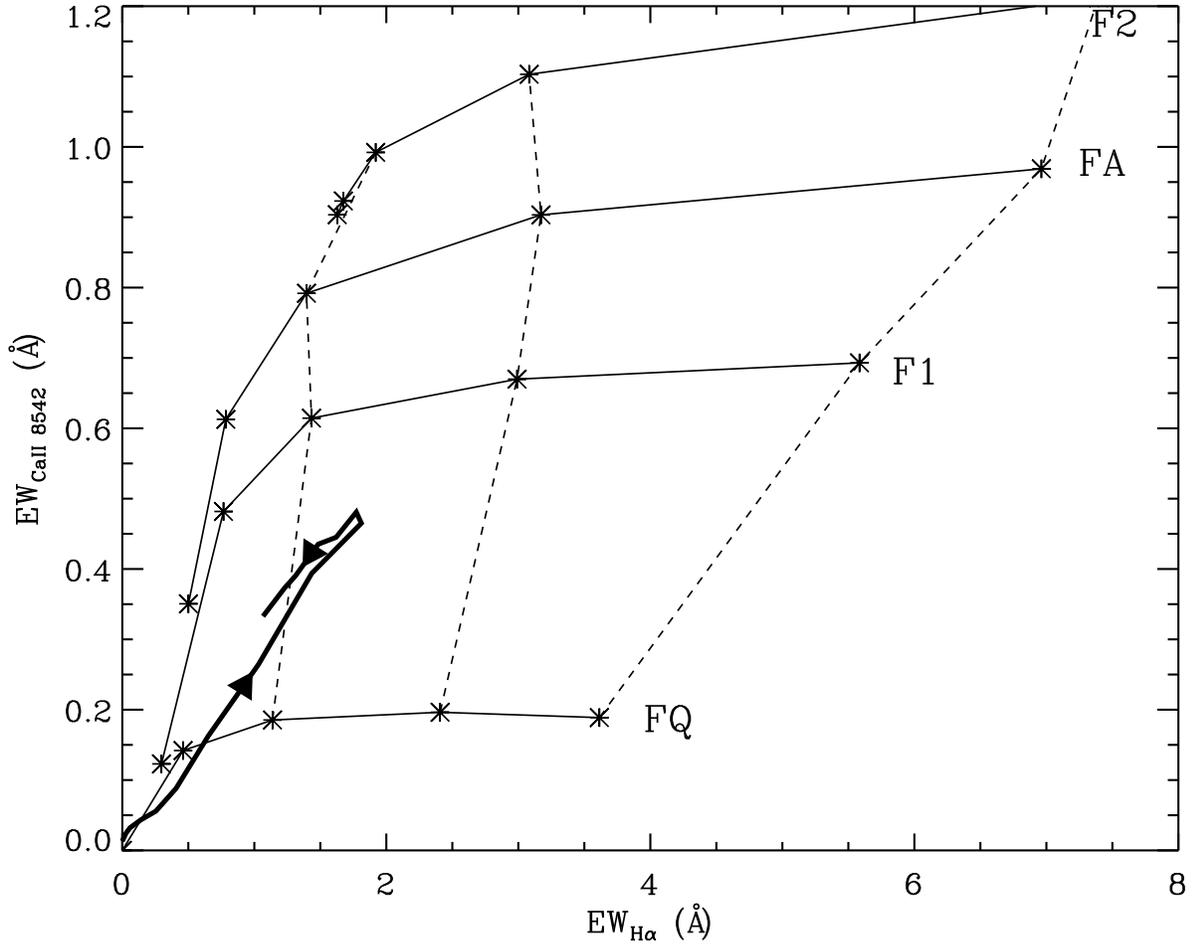} \caption{Observed Ca {\sc ii} 8542
{\AA} versus H$\alpha$ equivalent widths, showing the temporal
evolution of kernel K2, superimposed on the theoretical EW-EW
plot. A time range from 00:51:26 UT to 00:56:18 UT is represented
by arrows.}
\end{figure}
\begin{figure}
\epsscale{1.0} \plotone{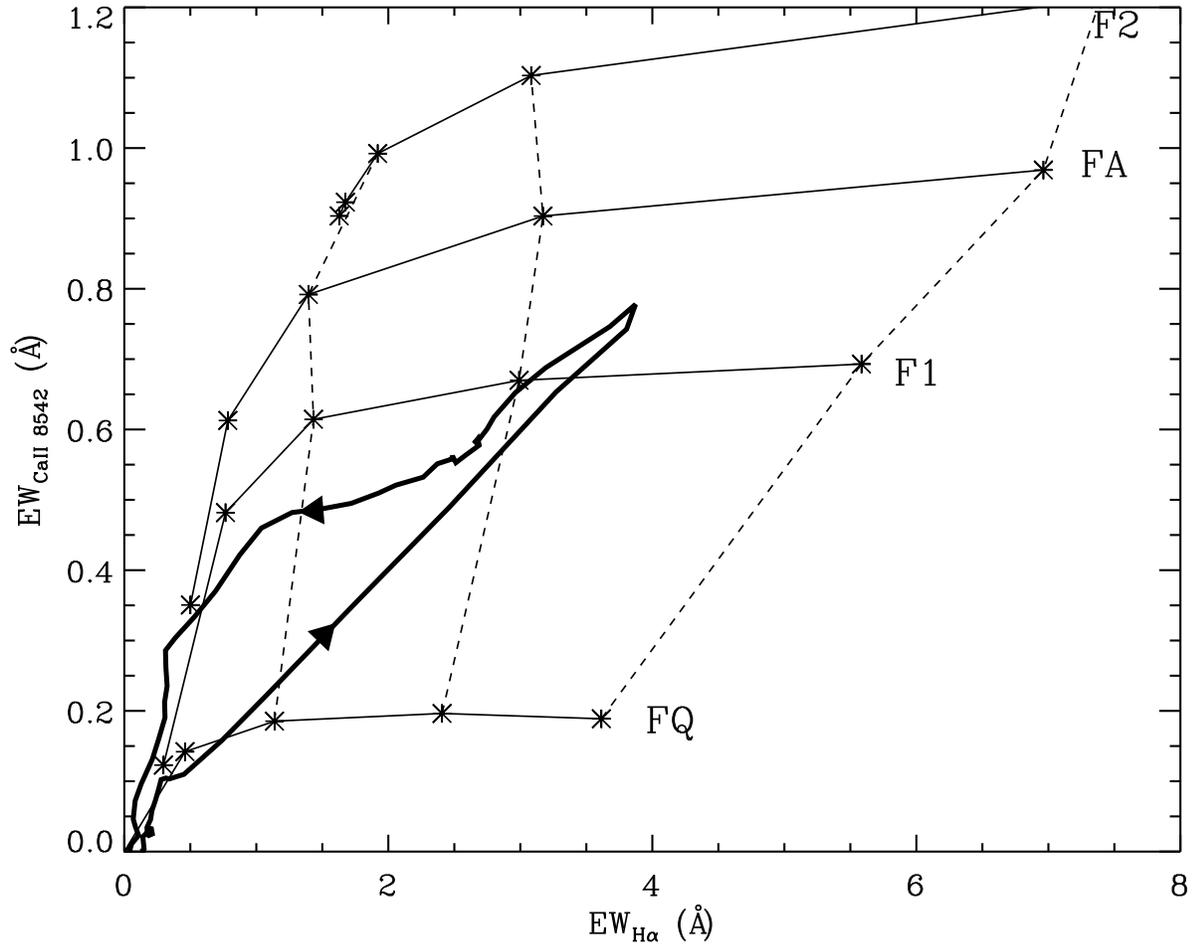} \caption{Same as Fig. 6, but for
kernel K3 with a time range from 00:53:49 UT to 01:45:56 UT.}
\end{figure}
\begin{figure}
\epsscale{1.0} \plotone{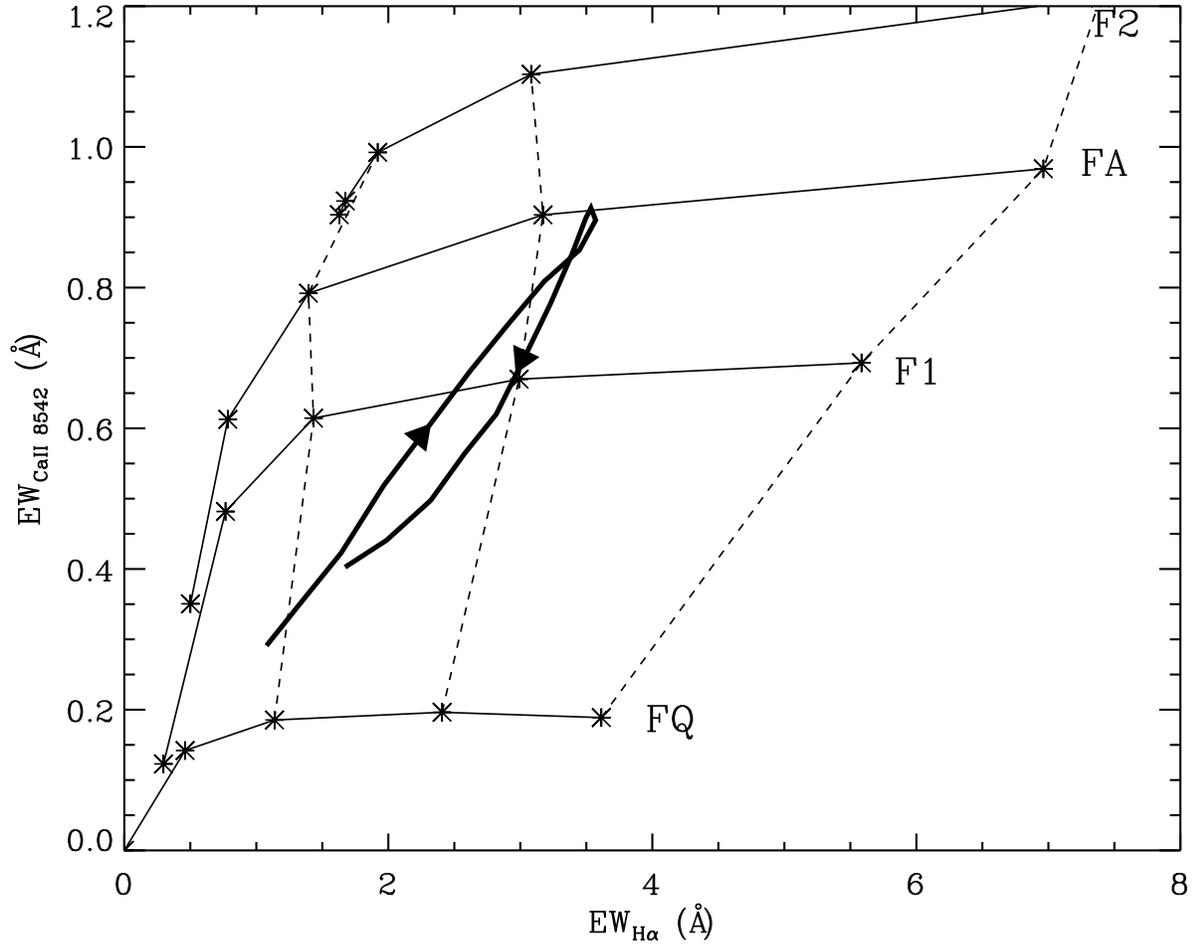} \caption{Observed Ca {\sc ii} 8542
{\AA} versus H$\alpha$ equivalent widths showing the spatial
variation along the line (Fig. 5), superimposed on the theoretical
EW-EW plot. From inner to outer edges across the flare ribbon, the
spatial variation is represented by arrows.}
\end{figure}

\end{document}